\documentclass[reprint,noshowpacs,noshowkeys,prd,balancelastpage,nofootinbib]{revtex4}
\usepackage[utf8]{inputenc}
\usepackage{color}
\usepackage{amsmath}
\usepackage{amssymb}
\usepackage{graphicx}
\usepackage[unicode=true,
 bookmarks=false,
 breaklinks=false,pdfborder={0 0 1},backref=section,colorlinks=true]{hyperref}
\usepackage{amsfonts}
\usepackage{mdframed}
\usepackage{footnote}
\usepackage{float}
\usepackage[font=footnotesize,it]{caption}
\usepackage{tikz}
\usepackage{tikz-3dplot}

\hypersetup{
 linkcolor=blue,urlcolor=blue,citecolor=red}
\makeatletter
\@ifundefined{textcolor}{}
{
 \definecolor{BLACK}{gray}{0}
 \definecolor{WHITE}{gray}{1}
 \definecolor{RED}{rgb}{1,0,0}
 \definecolor{GREEN}{rgb}{0,1,0}
 \definecolor{BLUE}{rgb}{0,0,1}
 \definecolor{CYAN}{cmyk}{1,0,0,0}
 \definecolor{MAGENTA}{cmyk}{0,1,0,0}
 \definecolor{YELLOW}{cmyk}{0,0,1,0}
}
\makeatother

\begin{document}

\title{Black-hole solution in nonlinear electrodynamics with the maximum
allowable symmetries}
\author{Z. Amirabi}
\email{zahra.amirabi@emu.edu.tr}
\author{S. Habib Mazharimousavi}
\email{habib.mazhari@emu.edu.tr}
\affiliation{Department of Physics, Faculty of Arts and Sciences, Eastern Mediterranean
University, Famagusta, North Cyprus via Mersin 10, Turkey}
\date{\today }

\begin{abstract}
The nonlinear Maxwell Lagrangian preserving both conformal and SO(2)
duality-rotation invariance has been introduced very recently. Here, in the
context of Einstein's theory of gravity minimally coupled with this
nonlinear electrodynamics, we obtain a black hole solution which is the
Reissner-Nordstr\"{o}m black hole with one additional parameter that is
coming from the nonlinear theory. We employ the causality and unitarity
principles to identify an upper bound for this free parameter. The effects
of this parameter on the physical properties of the black hole solution are
investigated.
\end{abstract}

\maketitle

\section{Introduction}

There are different models for nonlinear electrodynamics. The first such
model, known as Born-Infeld (BI) nonlinear electrodynamics, which is fully
relativistic and gauge-invariant, was proposed by Max Born and Leopold
Infeld in 1934 \cite{BI-1}. The initial idea was to modify Maxwell's linear
Lagrangian i.e., $\mathcal{L}=-F_{\mu \nu }F^{\mu \nu }$ to construct a
nonlinear Lagrangian with respect to Maxwell's invariants $\mathcal{S}%
=F_{\mu \nu }F^{\mu \nu }$ and $\mathcal{P}=F_{\mu \nu }\tilde{F}^{\mu \nu }$
such that the self-energy and the fields of a point charge remain finite at
the location of the charge. Furthermore, the vacuum polarization phenomena
in quantum electrodynamics (QED) has been observed experimentally since the
1940s. It is the polarization of virtual electron-positron pairs in vacuum
that is an indication for the nonlinear interaction of electromagnetic
fields such as photon-photon scattering. The interaction between photons can
be explained using the so-called Heisenburg-Euler (HE) effective-field
theory. The HE model was proposed by W. Heisenburg and H. Euler in 1936 \cite%
{EH} and is valid in the weak-field limit and large wavelengths. There are
other nonlinear electrodynamic models that have been introduced more
recently. For instance, the Logarithmic \cite{Log}, the Maxwell Power Law 
\cite{MPL,MPL1}, the arcsin \cite{Arcsin}, the rational \cite{Rational}, the
exponential \cite{Exp} and the double-Logarithmic \cite{DLog} models are
among them which all reproduce Maxwell's linear model in the weak-field
limit except the Maxwell Power Law. Furthermore, there are NED models that
don't reduce to the linear one in the weak-filed limit. Such models have
been coupled to Einstein's theory for constructing regular electric black
holes \cite{AyB,AyB2}. As it was proved by Bronnikov \cite{Bronnikov},
unlike the existence of a regular magnetic black hole, a regular electric
black hole solution doesn't exist in the gravity coupled with a NED which
yields Maxwell's theory in the weak-field limit.

In general, a generic NED model doesn't admit the symmetries of Maxwell's
theory. Among them are preserving conformal and SO(2) duality-rotation
invariance symmetries. In Ref. \cite{Conformal} a NED model has been
introduced which respects these symmetries (see Eq. (\ref{eq:Lagrangian})
below). In this interesting model, there is also a constant $\gamma ,$ which
is, in accordance with \cite{Conformal} and \cite{Kosyakov2}, a positive
parameter. In this study, we would like to apply the so-called causality and
unitarity principles for making an estimation for the upper bound of the
parameter $\gamma $. We would also like to examine the effects of this
parameter in the physical properties of the black hole solution in the
context of gravity coupled with this specific NED model. Since the black
hole is a dyonic solution, it is worth mentioning that such solutions have
already been found in the literature. In \cite{String}, dyonic black holes
(DBH) are found in string theory and in \cite{GR} DBH is found in gravity
rainbow. Also, DBH in dilatonic gravity and in nonlinear electrodynamics
coupled with gravity have been introduced in \cite{DG} and \cite{NLED},
respectively.

Finally, we would like to add that it is the conformal invariant symmetry of
the Maxwell theory which results in a traceless energy-momentum tensor i.e., 
$T_{\mu }^{\mu }=0.$ The same symmetry in a NED field theory also yields a
traceless energy-momentum tensor. This fact has been studied in \cite{MPL}
as well as in \cite{ConfMaxwell,ConfMaxwell2}.

Our Letter is organized as follows. In Sec. II we present the NED model that
admits conformal and SO(2) duality-rotation invariance symmetries. In Sec.
III we find the black hole solution of the gravity minimally coupled with
this NED. In Sec. IV we study the thermal stability of the solution. We
conclude our work in Sec. V.

\section{The Model}

The nonlinear Maxwell's Lagrangian is given by 
\begin{equation}
\mathcal{L}\left( \mathcal{S},\mathcal{P}\right) =-\mathcal{S}\cosh \gamma +%
\sqrt{\mathcal{S}^{2}+\mathcal{P}^{2}}\sinh \gamma  \label{eq:Lagrangian}
\end{equation}%
which has been first proposed in \cite{Conformal} and then re-proposed in 
\cite{Kosyakov2}. Considering the electromagnetic two-form, given by 
\begin{equation}
\mathbf{F}=\frac{1}{2}F_{\mu \nu }dx^{\mu }\wedge dx^{\nu }
\end{equation}%
in which 
\begin{equation}
F_{\mu \nu }=\partial _{\mu }A_{\nu }-\partial _{\nu }A_{\mu }
\end{equation}%
is the electromagnetic field tensor and 
\begin{equation}
\mathbf{A}=A_{\mu }dx^{\mu }
\end{equation}%
is the gauge potential one-form, the Maxwell invariants are defined to be $%
\mathcal{S}=F_{\mu \nu }F^{\mu \nu }$ and $\mathcal{P}=F_{\mu \nu }\tilde{F}%
^{\mu \nu }$ where 
\begin{equation}
\mathbf{\tilde{F}}=\frac{1}{2}\tilde{F}_{\mu \nu }dx^{\mu }\wedge dx^{\nu }
\end{equation}%
is the Hodge dual two-form of $\mathbf{F}$ with $\tilde{F}^{\mu \nu }=\frac{1%
}{2}\epsilon ^{\mu \nu \alpha \beta }F_{\alpha \beta }.$ In accordance with 
\cite{Conformal} and \cite{Kosyakov2}, $\gamma $ is a positive parameter,
however, we would like to see its possible upper bound by applying the
causality and unitarity conditions. Under the causality principle, the group
velocity of the elementary electromagnetic excitations should be less than
the speed of light in the vacuum and therefore there will be no tachyons in
the theory spectrum. Also, the unitarity principle requires the positive
definiteness of the norm of every elementary excitation of the vacuum upon
which ghosts are avoided. To obtain the necessary conditions imposed on any
NED due to the casualty and unitarity principles, basically one should study
the propagation of an electromagnetic wave in a spacetime filled with a
background electromagnetic field that is constant in time and space. In Ref. 
\cite{UnitCas1}, the corresponding dispersion relation for a general NED
Lagrangian has been found. In Ref. \cite{UnitCas}, a simplified version of
the former reference has been considered where the background
electromagnetic field was either purely electric or purely magnetic with $%
\mathcal{P}=0$. In this configuration, due to the phenomenon known as
birefringence, the propagating electromagnetic wave splits into two
orthogonal propagating modes. The requirement constraints found in \cite%
{UnitCas}, are applied to each of these modes and are given in terms of some
inequality relations as 
\begin{equation}
\mathcal{L}_{\mathcal{S}}\leq 0,\mathcal{L}_{\mathcal{SS}}\geq 0,\mathcal{L}%
_{\mathcal{PP}}\geq 0  \label{6}
\end{equation}%
and 
\begin{equation}
\mathcal{L}_{\mathcal{S}}+2\mathcal{SL}_{\mathcal{SS}}\leq 0,2\mathcal{SL}_{%
\mathcal{PP}}-\mathcal{L}_{\mathcal{S}}\geq 0.  \label{7}
\end{equation}%
Redefining $\mathcal{L}\left( \mathcal{S},\mathcal{P}\right) =-\mathcal{S}%
y\left( z\right) $ with 
\begin{equation}
y\left( z\right) =\cosh \gamma -sgn\left( \mathcal{S}\right) \sqrt{1+z^{2}}%
\sinh \gamma
\end{equation}%
and $z=\frac{\mathcal{P}}{\mathcal{S}},$ these inequalities reduce to 
\begin{equation}
y-zy^{\prime }\geq 2y^{\prime \prime }\geq 0\text{ }  \label{9}
\end{equation}%
for $\mathcal{S}<0$ and 
\begin{equation}
y-zy^{\prime }\geq -2z^{2}y^{\prime \prime }\geq 0  \label{10}
\end{equation}%
for $\mathcal{S}>0.$ Considering the explicit form of $y\left( z\right) $ we
find 
\begin{equation}
y^{\prime }=-sgn\left( \mathcal{S}\right) \frac{z\sinh \gamma }{\sqrt{1+z^{2}%
}}
\end{equation}%
and 
\begin{equation}
y^{\prime \prime }=-sgn\left( \mathcal{S}\right) \frac{\sinh \gamma }{\left(
1+z^{2}\right) ^{3/2}}.
\end{equation}%
Clearly, with $\gamma >0,$ $y^{\prime \prime }$ is definite-positive and
definite-negative for $\mathcal{S}<0$ and $\mathcal{S}>0,$ respectively.
Hence, (9) and (10) reduce to $y-zy^{\prime }-2y^{\prime \prime }\geq 0$ for 
$\mathcal{S}<0$ and $y-zy^{\prime }+2z^{2}y^{\prime \prime }\geq 0$ for $%
\mathcal{S}>0,$ respectively. In Fig. 1 we plot $K=y-zy^{\prime
}+2z^{2}y^{\prime \prime }$ in terms of $z$ for different values of $\gamma
. $ Our numerical calculation shows that for $\mathcal{S}>0$, (\ref{10}) is
satisfied provided $0<\gamma <\gamma _{\max }=\tanh ^{-1}\left( \frac{\sqrt{2%
}}{2}\right) .$ A similar numerical calculation reveals that for $\mathcal{S}%
<0$, (\ref{9}) is satisfied provided $0<\gamma <\infty .$ Therefore, in
order to satisfy all conditions with $\mathcal{S}>0$ and $\mathcal{S}<0$, we
impose $0<\gamma <\gamma _{\max },$ which is the intersection of the two
individual intervals. It is worth mentioning that for systems with no
magnetic charge/field such as the Hydrogen atom, $\mathcal{P}=0$ upon which
the Lagrangian reduces to the linear Maxwell's theory provided $\gamma =0$.

Finally, at the end of this section, we conclude that $\gamma $ which is a
dimensionless parameter of the theory has to be bounded from above as well
as from below i.e., $0<\gamma <\gamma _{\max }$. Therefore, through the rest
of the paper, we shall consider $\gamma $ to be in this interval.

\begin{figure}[tbp]
\caption{\protect\includegraphics[scale=0.6]{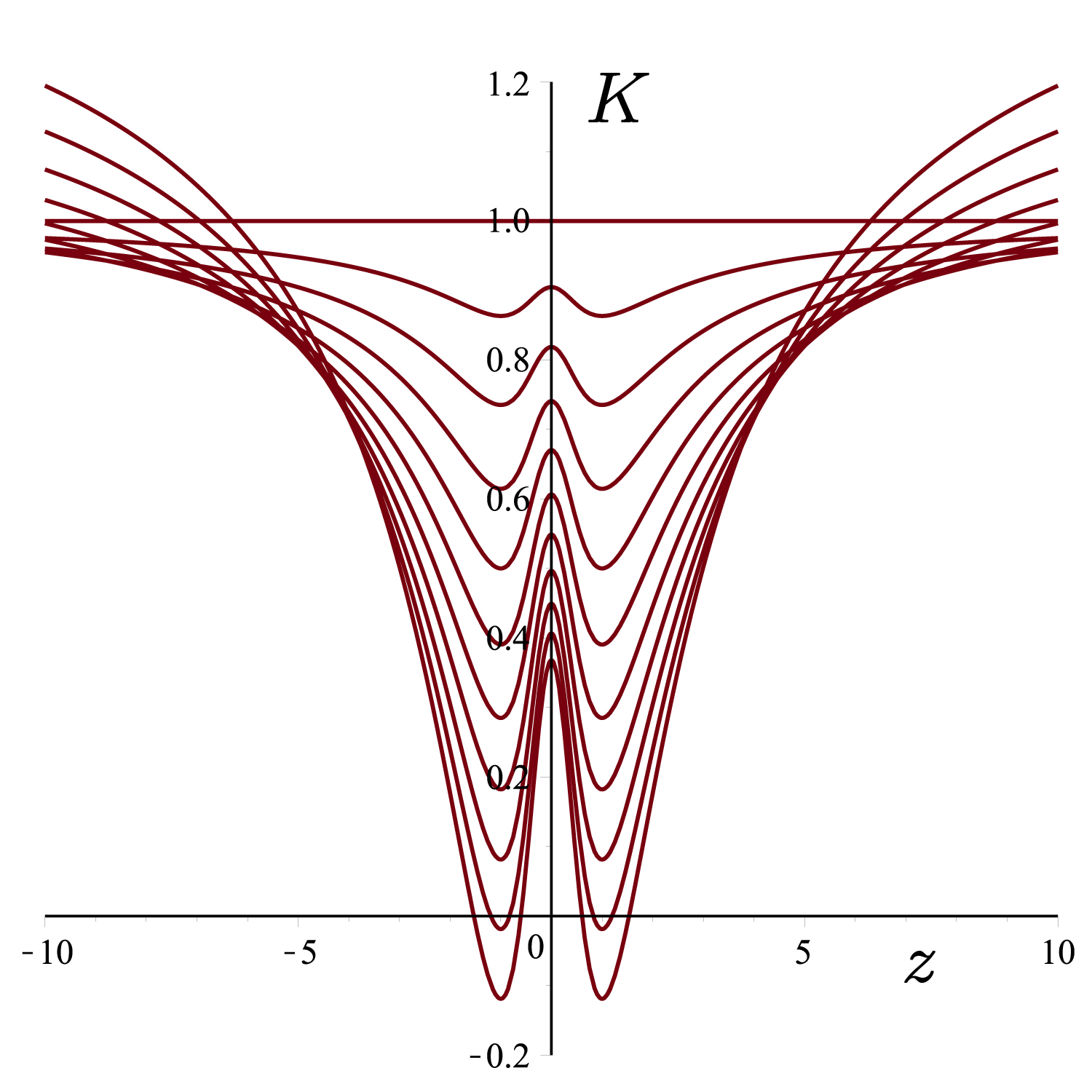}}
%\caption{\protect\includegraphics[scale=0.6]{Fig01}}
Plot of $K=y-zy^{\prime }+2z^{2}y^{\prime \prime }$ in terms of $z=\frac{%
\mathcal{P}}{\mathcal{S}}$ for $\gamma =0.0000$ to $\gamma =1.0000$ with
equal steps ($=0.1000$) from top to bottom. $K$ remains positive as long as $%
\gamma <\tanh ^{-1}\left( \frac{\sqrt{2}}{2}\right) .$%
\end{figure}

\section{The field equations and the black hole solution}

The action of Einstein-nonlinear-Maxwell theory is given by ($8\pi G=1$) 
\begin{equation}
I=\frac{1}{2}\int \sqrt{-g}d^{4}x\left( \mathcal{R}+\mathcal{L}\left( 
\mathcal{S},\mathcal{P}\right) \right)
\end{equation}%
in which $\mathcal{L}\left( \mathcal{S},\mathcal{P}\right) $ is given by Eq.
(\ref{eq:Lagrangian}). Upon applying the causality and unitarity conditions
we have already obtained an upper limit for $\gamma$ i.e., $\gamma <\tanh
^{-1}\left( \frac{\sqrt{2}}{2}\right) .$ Moreover, 
\begin{equation}
\lim_{\gamma \rightarrow 0}\mathcal{L}\left( \mathcal{S},\mathcal{P}\right)
=-\mathcal{S},
\end{equation}%
which is the linear Maxwell theory, however, it isn't the weak-field limit
of the Lagrangian (\ref{eq:Lagrangian}). The static spherically symmetric
spacetime and the electromagnetic two-form are chosen to be 
\begin{equation}
ds^{2}=-\psi \left( r\right) dt^{2}+\frac{dr^{2}}{\psi \left( r\right) }%
+r^{2}\left( d\theta ^{2}+\sin ^{2}\theta d\phi ^{2}\right)
\end{equation}%
and 
\begin{equation}
\mathbf{F}=Edt\wedge dr+Br^{2}\sin \theta d\theta \wedge d\phi ,  \label{EMF}
\end{equation}%
respectively, in which $E$ and $B$ are the radial components of the static
electric and magnetic fields indicating the presence of the electric and
magnetic monopoles. Variation of the action with respect to the metric
tensor implies the Einstein-nonlinear Maxwell equations given by 
\begin{equation}
G_{\mu }^{\nu }=T_{\mu }^{\nu }
\end{equation}%
in which 
\begin{equation}
T_{\mu }^{\nu }=\frac{1}{2}\left( \left( \mathcal{L}-\mathcal{PL}_{\mathcal{P%
}}\right) \delta _{\mu }^{\nu }-4\mathcal{L}_{\mathcal{S}}F_{\mu \lambda
}F^{\nu \lambda }\right)
\end{equation}%
is the energy-momentum tensor and $G_{\mu }^{\nu }$ is the standard Einstein's
tensor. We note that, $\mathcal{L}_{\mathcal{S}}=\frac{\partial \mathcal{L}}{%
\partial \mathcal{S}}$ and $\mathcal{L}_{\mathcal{P}}=\frac{\partial 
\mathcal{L}}{\partial \mathcal{P}}.$ Furthermore, the variation of the
action with respect to the four-potential yields the Maxell-nonlinear
equations 
\begin{equation}
d\left( \mathcal{L}_{\mathcal{S}}\mathbf{\tilde{F}}+\mathcal{L}_{\mathcal{P}}%
\mathbf{F}\right) =0  \label{MNLE}
\end{equation}%
where $\mathbf{\tilde{F}}$ is the dual two-form of $\mathbf{F}$ which is
found to be 
\begin{equation}
\mathbf{\tilde{F}=-}Bdt\wedge dr+Er^{2}\sin \theta d\theta \wedge d\phi .
\label{EMFd}
\end{equation}%
Having, $\mathbf{F}$ and $\mathbf{\tilde{F}}$ given by (\ref{EMF}) and (\ref%
{EMFd}) we obtain 
\begin{equation}
\mathcal{S}=2\left( B^{2}-E^{2}\right)
\end{equation}%
and 
\begin{equation}
\mathcal{P}=4EB,
\end{equation}%
upon which, the Maxell-nonlinear equations (\ref{MNLE}) reduce to the
following two individual equations 
\begin{equation}
d\left( \left( -\mathcal{L}_{\mathcal{S}}B+\mathcal{L}_{\mathcal{P}}E\right)
dt\wedge dr\right) =0  \label{ME1}
\end{equation}%
and 
\begin{equation}
d\left( \left( \mathcal{L}_{\mathcal{S}}E+\mathcal{L}_{\mathcal{P}}B\right)
r^{2}\sin \theta d\theta \wedge d\phi \right) =0.  \label{ME2}
\end{equation}%
From the Bianchi identity, i.e., 
\begin{equation}
d\mathbf{F=0}
\end{equation}%
which implies 
\begin{equation*}
d\left( Edt\wedge dr+Br^{2}\sin \theta d\theta \wedge d\phi \right) =0,
\end{equation*}%
one finds that both radial fields i.e., $E$ and $B,$ and consequently the
invariants $\mathcal{S}$ and $\mathcal{P}$ should be only functions of $r$.
Hence, (\ref{ME1}) is trivially satisfied and (\ref{ME2}) suggests 
\begin{equation}
\left( \mathcal{L}_{\mathcal{S}}E+\mathcal{L}_{\mathcal{P}}B\right) r^{2}=c
\label{MFE}
\end{equation}%
in which $c$ is an integration constant. Furthermore, the Bianchi identity
implies that, 
\begin{equation}
B=\frac{Q_{m}}{r^{2}}  \label{MagneticField}
\end{equation}%
where $Q_{m}$ is an integration constant representing the magnetic charge.
Considering, (\ref{MagneticField}) and (\ref{MFE}) together with the
Maxwell's invariants, one obtains 
\begin{equation}
E=\frac{Q_{e}}{r^{2}}
\end{equation}%
in which $Q_{e}$ is a constant representing the electric charge,
satisfying 
\begin{equation}
\left( zQ_{e}-Q_{m}\right) y^{\prime }-Q_{e}y=c
\end{equation}%
with 
\begin{equation}
z=\frac{\mathcal{P}}{\mathcal{S}}=\frac{2Q_{e}Q_{m}}{Q_{m}^{2}-Q_{e}^{2}}
\end{equation}%
which is a constant. The explicit form of Maxwell's invariants are given by 
\begin{equation}
\mathcal{S}=2\frac{Q_{m}^{2}-Q_{e}^{2}}{r^{4}}
\end{equation}%
and 
\begin{equation}
\mathcal{P}=\frac{4Q_{m}Q_{e}}{r^{4}}.
\end{equation}%
Following the nonlinear-Maxwell equations, we shall solve the
Einstein-nonlinear Maxwell equations. To do so, we find the nonzero
components of the energy momentum-tensor given by 
\begin{equation}
T_{t}^{t}=T_{r}^{r}=\frac{1}{2}\left[ \left( \mathcal{L}-\mathcal{PL}_{%
\mathcal{P}}\right) +4\mathcal{L}_{\mathcal{S}}E^{2}\right] =\left(
y^{\prime }z-y\right) \frac{Q_{m}^{2}+Q_{e}^{2}}{r^{4}}
\end{equation}%
and 
\begin{equation}
T_{\theta }^{\theta }=T_{\phi }^{\phi }=\frac{1}{2}\left[ \left( \mathcal{L}-%
\mathcal{PL}_{\mathcal{P}}\right) -4\mathcal{L}_{\mathcal{S}}B^{2}\right]
=-\left( y^{\prime }z-y\right) \frac{Q_{m}^{2}+Q_{e}^{2}}{r^{4}}.
\end{equation}%
Using a fluid model for the energy momentum tensor i.e., $T_{\mu }^{\nu
}=diag\left( -\rho ,p_{r},p_{\theta },p_{\phi }\right) $ one finds 
\begin{equation}
\rho =\left( y-zy^{\prime }\right) \frac{Q_{m}^{2}+Q_{e}^{2}}{r^{4}}%
=-p_{r}=p_{\theta }=p_{\phi }.
\end{equation}%
Having, 
\begin{equation}
y-zy^{\prime }=\cosh \gamma \left( 1-\frac{sgn\left( \mathcal{S}\right)
\tanh \gamma }{\sqrt{1+z^{2}}}\right) ,
\end{equation}%
which is definite-positive for all values of $0<\gamma <\gamma _{\max },$ $%
\mathcal{S}$ and $z,$ we obtain $\rho \geq 0$ and $\rho +p_{i}\geq 0$ which
in turn imply that the weak energy conditions are satisfied. Furthermore,
the strong energy conditions i.e., $\rho +p_{i}\geq 0$ $\ $and $\rho
+\sum_{i}p_{i}\geq 0$ are also satisfied.

Next, we introduce 
\begin{equation}
\omega ^{2}=y-zy^{\prime }=\cosh \gamma \left( 1-\frac{1-q^{2}}{1+q^{2}}%
\tanh \gamma \right) ,
\end{equation}%
which is definite-positive with $q^{2}=\frac{Q_{e}^{2}}{Q_{m}^{2}}$. This is
because of the causality and unitarity conditions upon which we imposed $%
0<\gamma <\gamma _{\max }$. Hence, the energy momentum-tensor simplifies as 
\begin{equation}
T_{\mu }^{\nu }=\omega ^{2}\frac{Q_{m}^{2}+Q_{e}^{2}}{r^{4}}diag\left(
-1,-1,1,1\right) .
\end{equation}%
Finally, the Einstein-nonlinear-Maxwell equations admit 
\begin{equation}
\psi \left( r\right) =1-\frac{2M}{r}+\omega ^{2}\frac{Q_{m}^{2}+Q_{e}^{2}}{%
r^{2}}  \label{BHS}
\end{equation}%
in which $M$ is an integration constant, representing the mass of the black
hole. This is a dyonic Reissner-Nordstr\"{o}m-type \cite{RN} charged black
hole solution with an additional parameter $\gamma .$ Hence, the general
properties of (\ref{BHS}), are similar to RN black hole. In the next
section, we study the effects of the parameter $\gamma $ in the thermal
stability of the black hole. For the thermodynamic theory of black holes
including RN, we refer to \cite{Davis} while for the phase transition in RN
black hole we refer to \cite{PTr}. 
\begin{figure}[tbp]
\caption{\protect\includegraphics[scale=0.6]{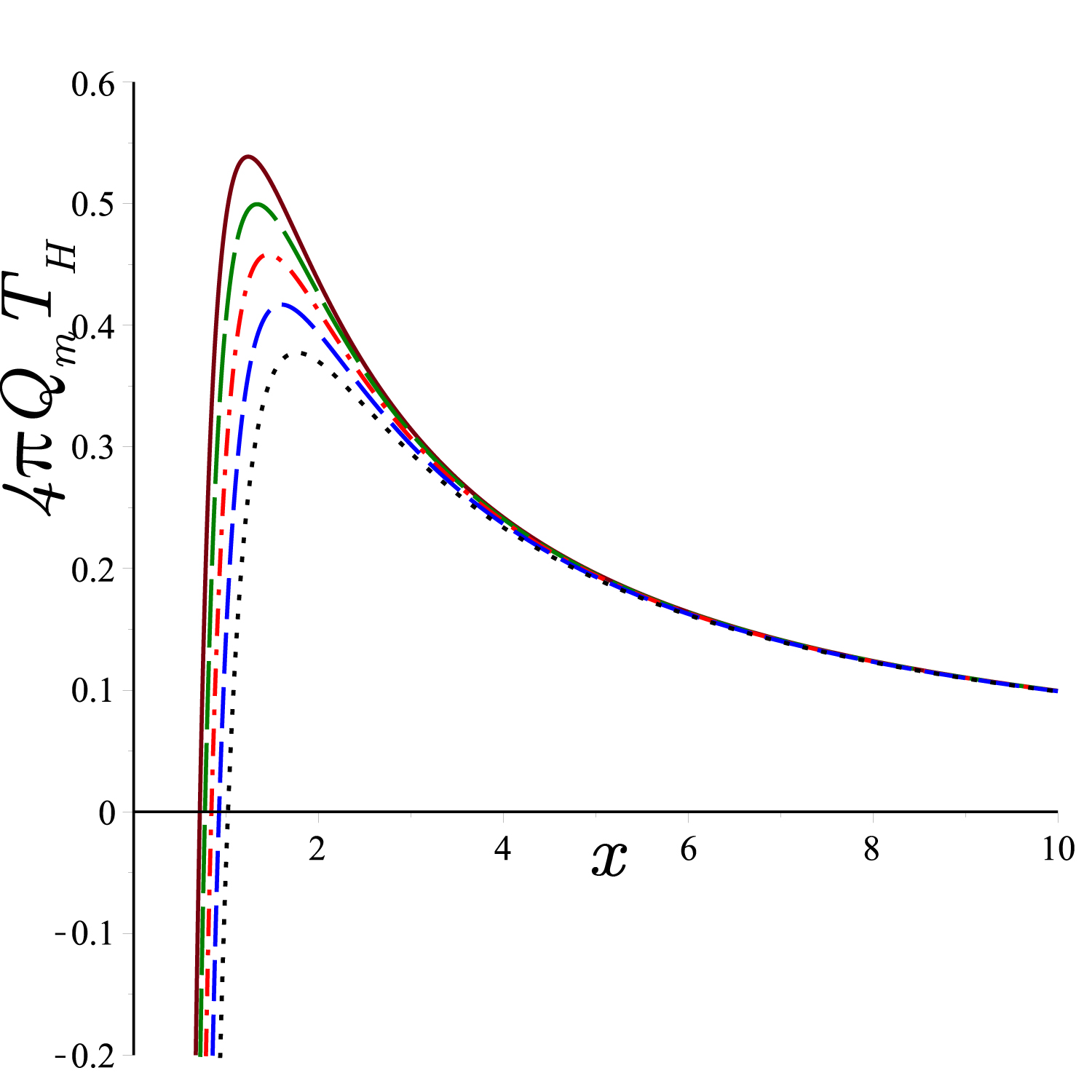}}
%\caption{\protect\includegraphics[scale=0.6]{Fig02}}
The Hawking temperature $4\pi Q_{m}T_{H}$ in terms of $x=\frac{r_{h}}{Q_{m}}$
for $\gamma =0.0000$ to $\gamma =\tanh ^{-1}\left( \frac{\sqrt{2}}{2}\right) 
$ with equal steps (from bottom to top) and $\left( \frac{Q_{e}}{Q_{m}}%
\right) =0.2.$%
\end{figure}
\begin{figure}[tbp]
\caption{\protect\includegraphics[scale=0.6]{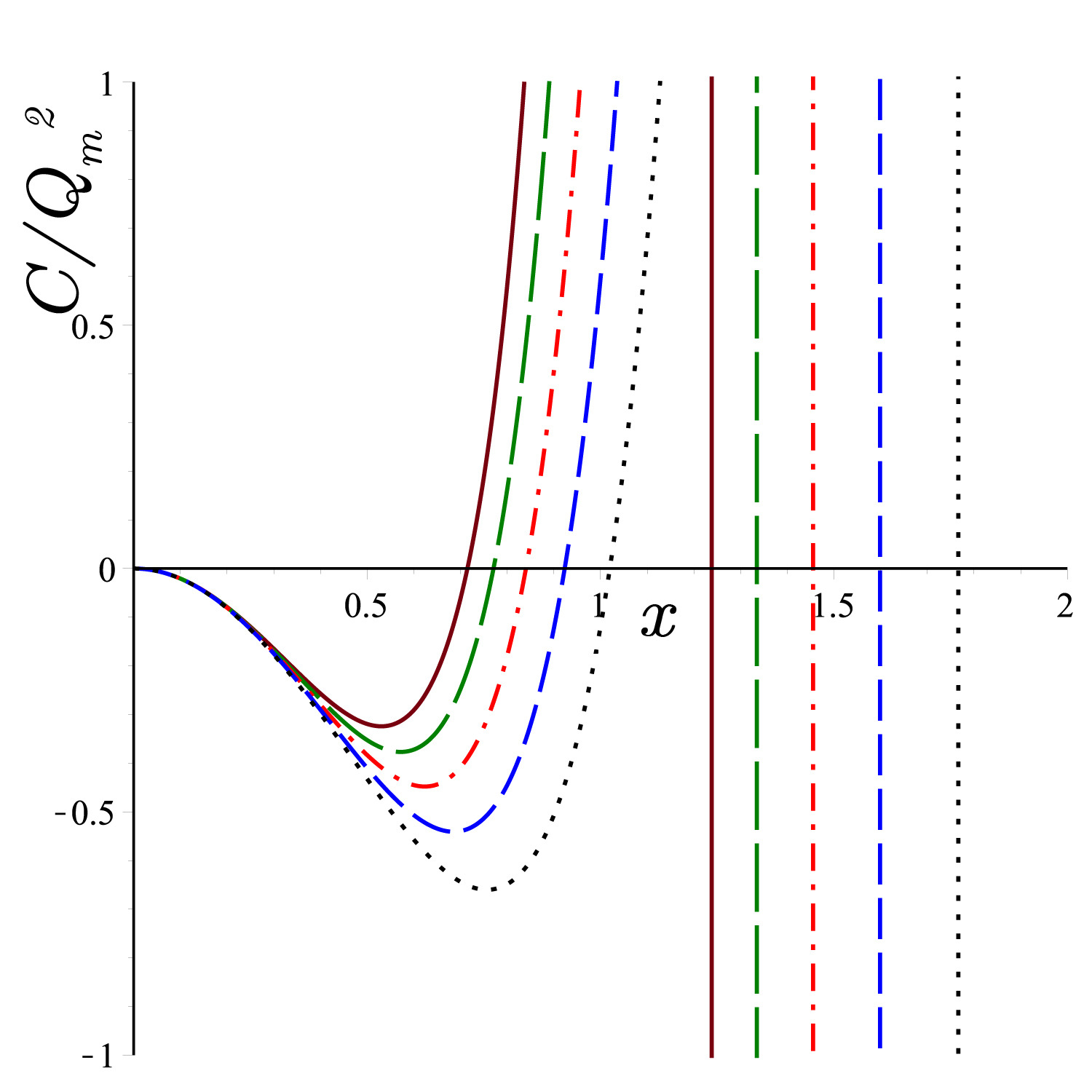}}
%\caption{\protect\includegraphics[scale=0.6]{Fig03}}
The heat capacity $C_{Q}/Q_{m}^{2}$ in terms of $x=\frac{r_{h}}{Q_{m}}$ for $%
\gamma =0.0000$ to $\gamma =\tanh ^{-1}\left( \frac{\sqrt{2}}{2}\right) $
with equal steps (from bottom to top) and $\left( \frac{Q_{e}}{Q_{m}}\right)
=0.2.$ Please note that, the second branch of the heat capacity is not
depicted.
\end{figure}

\section{Thermal Stability of the black hole solution}

To investigate the effects of the parameter $\gamma $ in the thermal
stability of the black hole solution (\ref{BHS}) we start with the Hawking
temperature which is given by 
\begin{equation}
T_{H}=\frac{\psi ^{\prime }\left( r_{h}\right) }{4\pi }=\frac{1}{4\pi r_{h}}%
\left( 1-\frac{\omega ^{2}Q^{2}}{r_{h}^{2}}\right)
\end{equation}%
in which $r_{h}$ is the radius of the event horizon and $%
Q^{2}=Q_{m}^{2}+Q_{e}^{2}$. In Fig. 2, we plot the $4\pi Q_{m}T_{H}$ versus $%
x=\frac{r_{h}}{Q_{m}}$ with $\frac{Q_{e}}{Q_{m}}=0.2$ and $\gamma =0$ to $%
\gamma =\gamma _{\max }$ with equal steps. Increasing the value of $\gamma ,$
for a given radius of the event horizon, increases the Hawking temperature.
Furthermore, the heat capacity for constant $Q$ is defined to be 
\begin{equation}
C_{Q}=\left( T_{H}\frac{\partial S}{\partial T_{H}}\right) _{Q}=-\frac{2\pi
r_{h}^{2}\left( r_{h}^{2}-\omega ^{2}Q^{2}\right) }{r_{h}^{2}-3\omega
^{2}Q^{2}}
\end{equation}%
where $S=\pi r_{h}^{2}$ is the entropy of the black hole. In Fig. 3 we plot $%
C_{Q}/Q_{m}^{2}$ with respect to $x=\frac{r_{h}}{Q_{m}}$ with $\frac{Q_{e}}{%
Q_{m}}=0.2$ and $\gamma =0$ to $\gamma =\gamma _{\max }$ with equal steps.
The Type-1 ($C_{Q}=0$) and Type-2 ($C_{Q}\rightarrow \pm \infty $)
transition points are emphasized. These points are given by 
\begin{equation}
\left( r_{h}\right) _{Type-1}=\omega Q
\end{equation}%
and 
\begin{equation}
\left( r_{h}\right) _{Type-2}=\sqrt{3}\omega Q.
\end{equation}%
Let's add that the thermal stability region is defined to admit both $T_{H}$
and $C_{Q}$ positive. Therefore, the black hole is thermally stable if $%
\left( r_{h}\right) _{Type-1}<r_{h}<\left( r_{h}\right) _{Type-2}.$ For the
specific value of $\frac{Q_{e}}{Q_{m}}=0.2$ it is observed from Fig. 3 that,
the transition points are shifted to the smaller values for the larger $\gamma $
which in turn yields a narrower stability region.
\begin{figure}[tbp]
\caption{\protect\includegraphics[scale=0.6]{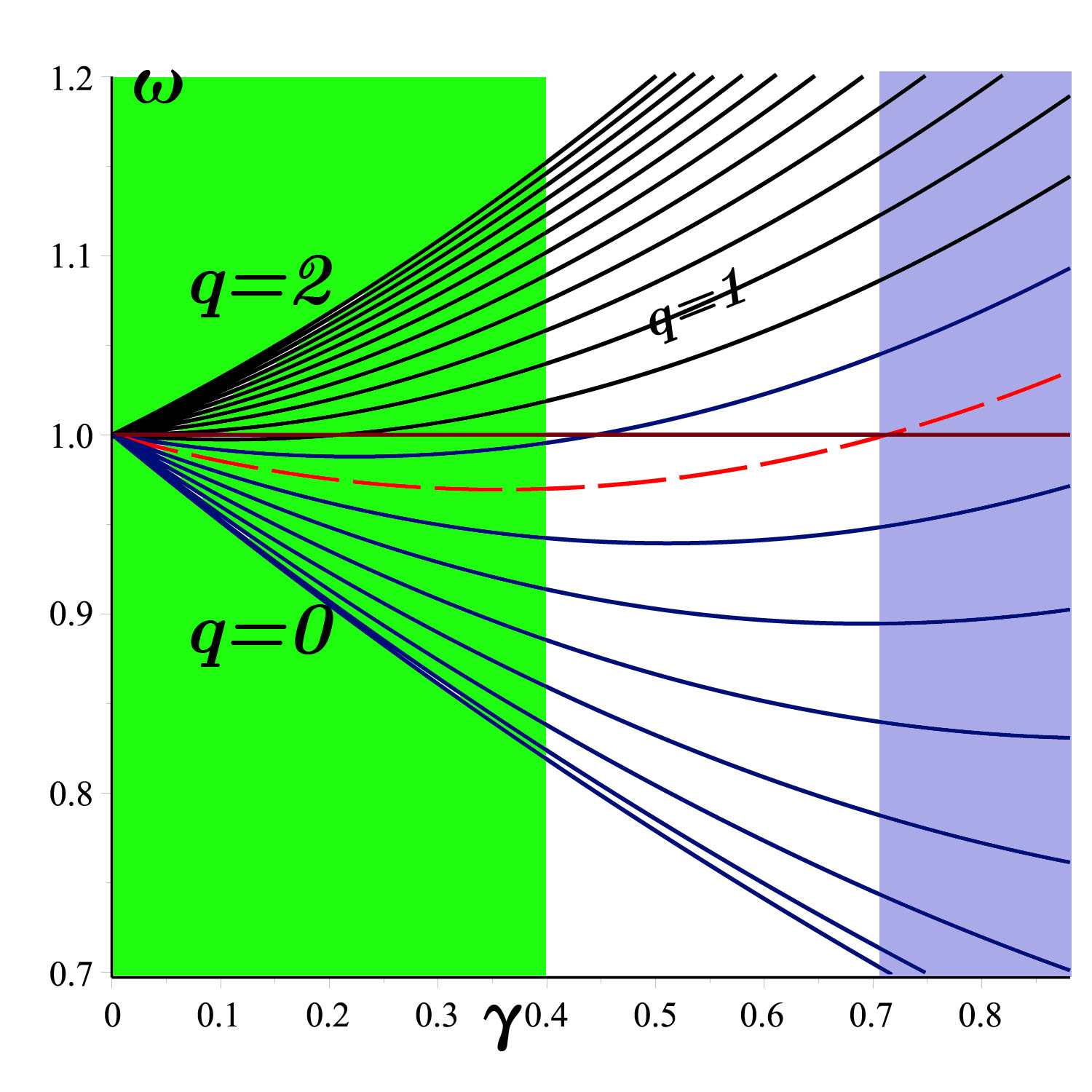}}
%\caption{\protect\includegraphics[scale=0.6]{Fig04}}
Plots of $\omega $ with respect to $\gamma $ for various value of $q=\frac{%
Q_{e}}{Q_{m}}=0...2$ with equal steps. The dashed curve is for the
particular $q=0.7$ and the three regions colored with green (left), white
(middle), and light-blue (right) imply $\omega $ decreasing and less than 1,
increasing and less than 1, and increasing and greater than 1, respectively.
For $q\geq 1,$ the curve\ of $\omega $ is increasing with respect to $\gamma 
$. These curves are depicted above the sign of $q=1.$%
\end{figure}
For the sake of completeness, we give a general overview of the stability
region. In Fig. 2 and 3, the value of $q$ was set to $0.2,$ however, for
larger $q$ the configuration changes. Let's define the width of the
stability region to be%
\begin{equation}
\triangle r_{h}=\left( r_{h}\right) _{Type-2}-\left( r_{h}\right)
_{Type-1}=\left( \sqrt{3}-1\right) \omega Q.
\end{equation}%
In Fig. 4 we plot $\omega $ versus $\gamma $ for the various value of $%
q=0...2$. It can be seen from Fig. 4 that the width of the region of
stability $\triangle r_{h}$ depends not only on $\gamma $ but also on $q.$
For $q=0.2,$ that we plot the corresponding $T_{H}$ and $C_{Q}$ in Fig. 2
and 3, $\omega $ is a decreasing function in the interval $0<\gamma <\gamma
_{\max }.$ Hence, we concluded that the region of stability decreases with
the increment of $\gamma .$ Our detailed calculation reveals that $\omega $
admits a minimum at $\gamma _{crit}=\ln \frac{1}{q}$ and becomes zero at $%
\gamma _{0}=2\gamma _{crit}.$ For $q<\frac{1}{1+\sqrt{2}}$, both $\gamma
_{crit}$ and $\gamma _{0}$ remain outside of the domain of $\gamma $ such
that with an increment in $\gamma $ the width of stability becomes smaller.
For $\frac{1}{1+\sqrt{2}}<q<\frac{1}{\sqrt{1+\sqrt{2}}},$ only $\gamma
_{crit}$ falls in the domain of $\gamma <\gamma _{\max }$ and consequently
the width of the stability region first decreases and then increases, even
though it remains less than the corresponding RN case. Finally, if $\frac{1}{%
\sqrt{1+\sqrt{2}}}<q<1$ then both $\gamma _{crit}$ and $\gamma _{0}$ remain
in the domain of acceptable $\gamma .$ Hence, $\triangle r_{h}$ first
decreases with $\gamma <$ $\gamma _{crit}$ then increases and remains less
than one with $\gamma <\gamma _{0}$ and finally increases to the values
greater than the corresponding RN case with $\gamma _{0}<\gamma <\gamma
_{\max }.$ In Fig. 4, these three regions of $\gamma $ are shown with
different shaded colors for a particular $q=0.7.$ Furthermore, for $q\geq 1,$
the graph of $\omega $ versus $\gamma $ is an increasing function, which
indicates that the width of the stability region increases. For this fact, we refer to the
curves after $q=1$ in Fig. 4.

\section{Conclusion}

We re-examined the recently introduced conformal and SO(2) duality-rotation
invariance NED model, given in Eq. (\ref{eq:Lagrangian}). The same model has
also been used in two very recent papers \cite{Danial} to study NUT
wormholes, Taub-Bolt instantons, black holes, and exact gravitational waves.
We applied the unitarity and casualty conditions to find an upper bound for
the arbitrarily dimensionless constant $\gamma $ in the theory. Following
our results, the domain of $\gamma$ has been found to be $0<\gamma <\gamma _{\max }=\tanh
^{-1}\left( \frac{\sqrt{2}}{2}\right) $. Furthermore, we minimally coupled
this particular NED with Einstein's gravity. From the field equations, we
obtained a Reissner-Nordstr\"{o}m-type charged black hole solution with a
new extra parameter, i.e., $\gamma $. Let's note that $\omega ^{2}=\cosh
\gamma -\frac{1-q^{2}}{1+q^{2}}\sinh \gamma $ represents $\gamma $ in our
investigation. The effects of $\gamma $ on the physical properties of the
black hole solutions have been investigated. The thermal stability of the
black hole, specifically, has been studied. The results have been
demonstrated in Fig. 2, Fig. 3, and Fig. 4. In accordance with our analysis,
for $0<q<1$ the stability region may increase or decrease depending on the
value of $q$ and $\gamma $. However, for $q\geq1,$ the stability region
increase with $\gamma $.

\end{document}